\documentclass[aps,prb,showpacs,reprint,groupedaddress]{revtex4-1}

\usepackage{graphicx}
\usepackage{epstopdf}

\begin{document}


\title{Quantum phenomena in the radial thermal expansion of bundles of single-walled carbon nanotubes doped with $^3$He. A giant isotope effect}


\author{A. V. Dolbin}
 \email{dolbin@ilt.kharkov.ua}
 \homepage{http://www.dolbin.org.ua/}
\author{V. B. Esel'son}
\author{V. G. Gavrilko}
\author{V. G. Manzhelii}
\author{N. A. Vinnikov}
\author{S. N. Popov}
\affiliation{B. Verkin Institute for Low Temperature Physics \& Engineering NASU, Kharkov 61103, Ukraine
}

\author{B. Sudqvist}
\affiliation{Department of Physics, Umea University, SE-901 87 Umea, Sweden
}

\date{\today}

\begin{abstract}
The radial thermal expansion $\alpha_r$ of bundles of single-walled carbon 
nanotubes saturated with $^3$He up to the molar concentration 9.4\% has been 
investigated in the temperature interval 2.1--9.5~K by high-sensitivity 
capacitance dilatometry. In the interval 2.1--7~K a negative $\alpha_r$ was 
observed, with a magnitude which exceeded the largest negative $\alpha_r$ 
values of pure and $^4$He-saturated nanotubes by three and two orders of 
magnitude, respectively. The contributions of the two He isotope impurities 
to the negative thermal expansion of the nanotube bundles are most likely 
connected with the spatial redistribution of $^4$He and $^3$He atoms by 
tunneling at the surface and inside nanotube bundles. The isotope 
effect turned out to be huge, probably owing to the higher tunneling 
probability of $^3$He atoms.
\end{abstract}



\maketitle

\section{Introduction}

Since their discovery by Iijima in 1991 \cite{Iijima-354-60-1991}, 
carbon nanotubes (CNTs)
 have been attracting intense interest from scientists owing to their unique
 geometry and extraordinary physical properties. The very high
 length-to-diameter ratios and the capability of CNTs to form bundles of
several tens or even hundreds of tubes make it possible to form
 low-dimensional, ordered impurity phases at the bundles' surfaces
\cite{Cole-84-3883-2000,Pearce-95-185302-2005}.
Such phases consist of impurity molecules or atoms forming one-dimensional
chains in the intertube grooves or in the interstitial channels in the bundles.
They can also form two-dimensional layers at the bundle surface. It has been
found experimentally \cite{Dolbin-35-613-2009,Dolbin-35-1209-2009,
Dolbin-36-465-2010,Dolbin-36-797-2010} that the radial thermal expansion coefficients
$\alpha_r$ of nanotube bundles are negative in the region of liquid helium
temperatures. Gas impurities usually suppress the magnitudes of these
negative values of $\alpha_r$ and reduce the temperature region where they
exist. The $^4$He impurity is an exception: when $^4$He is introduced both
the magnitude of the negative $\alpha_r$ values and the temperature region 
of the
negative thermal expansion increase \cite{Dolbin-36-797-2010}. This effect was attributed to a
tunneling redistribution of the $^4$He atoms at the surface and inside CNT
bundles. It is known \cite{Freiman-9-335-1983,Narayanamurti-42-201-1970}
 that the processes of tunneling gives a negative
contribution to the thermal expansion of a system. It is then reasonable
to expect that saturation of CNT bundles with $^3$He would enhance the above
effect because the smaller masses of $^3$He atoms must increase the
probability of tunneling.

In the present work we have, therefore, investigated the radial thermal
expansion of single-walled carbon nanotubes (SWNTs) saturated with $^3$He
using the dilatometric method. The temperature interval studied was
2.1--9.5~K. As will be shown below, the experimental results verify our
expectation that the addition of $^3$He should enhance the negative thermal
expansion; in fact the effect is surprisingly large, two orders of magnitude
larger than for $^4$He.

\section{Experimental technique}

The radial thermal expansion of $^3$He saturated CNTs was investigated using
a high-sensitivity low-temperature capacitance dilatometer with
$2\cdot10^{-9}$ cm resolution. The technique and the experimental apparatus are
presented in detail elsewhere \cite{Aleksandrovskii-23-1256-1997}. 
The sample was a cylinder 7.2 mm high
and 10 mm in diameter, obtained by compressing a stack of thin
($\le0.4$ mm) plates consisting of in-plane oriented CNTs at 1.1 GPa.
The plates were prepared by compressing (1.1 GPa) small amounts of CNT powder
(Cheap Tubes, USA, CCVD method). It is known \cite{Bendiab-93-1769-2002}
that such pressure treatment
of a thin CNT layer leads to a preferred orientation where the CNT axes mainly
lie in the plane perpendicular to the applied pressure, the average deviation
being about $4^\circ$. The alignment of CNT axes in the plane makes it
possible to investigate preferentially the radial component of the thermal
expansion of the tubes \cite{Dolbin-34-860-2008}
 and the effect of gas saturation upon the radial
thermal expansion of SWNT bundles \cite{Dolbin-35-613-2009,Dolbin-35-1209-2009,
Dolbin-36-465-2010,Dolbin-36-797-2010}.

Just before starting the investigation, the cell with a pure CNT sample was 
evacuated at room temperature for 72 hours to remove possible gas impurities. 
It was then cooled to 2.1~K and a series of control measurements was 
performed. The results showed that the thermal expansion of the sample 
coincided, within the experimental error, with the values obtained previously 
for pure CNTs \cite{Dolbin-34-860-2008} (see Fig. \ref{fig:1}a, curve 4). 
$^3$He gas was then fed to the measuring cell at T = 2.1~K. 
The $^3$He was added in small portions as some quantities were sorbed by the 
nanotubes. This permitted us to maintain the pressure in the cell several 
times lower than the pressure of saturated $^3$He vapor at this temperature 
(151.112 Torr at T = 2~K \cite{Sydoriak-68A-579-1964}). 
The total amount of $^3$He absorbed by 
the pure CNT sample was 9.4 mol. \% (94 $^3$He atoms per 1000 C atoms). 
At this impurity concentration we were able to compare our results on the 
thermal expansion of the $^3$He-SWNT with those from previous measurements 
of the radial thermal expansion of CNTs saturated with $^4$He to the molar 
concentration 9.4\% \cite{Dolbin-36-797-2010}. After the sorption was 
completed, an equilibrium of $\sim 1 \cdot 10^{-4}$ Torr was set in the 
measuring cell. Since the rise of the sample temperature in the course of 
measuring $\alpha_r$ could entail some $^3$He desorption from the sample, the 
reproducibility of the results was checked at regular intervals by heating 
and subsequently cooling the sample by $\Delta T$, 
where $\Delta T$ = 0.3 \ldots 1~K. 
If the results obtained under this cycling coincided, within experimental 
error, the effect of He desorption was regarded as negligible and the data 
were considered to be obtained in equilibrium. The absence of reproducibility 
was believed to show that at this and higher temperatures the desorption 
of the $^3$He impurity from the sample had some effect on the thermal 
expansion. The measurement was then stopped. Note that for the radial 
thermal expansion of the $^3$He-SWNT the data were observed to be 
reproducible in the interval T = 2.1--9.5 K, but no longer reproducible 
when cooling the sample to 9.7~K. When reproducibility was no longer 
observed the sample was heated to T = 11~K and held at this temperature 
under dynamic evacuation until an equilibrium pressure of $7.5\cdot10^{-2}$ 
Torr was achieved in the system. During this process, a fraction of the 
$^3$He impurity was desorbed from the sample. The sample was then cooled 
back down to the lowest temperature, 2.1~K, and the radial thermal 
expansion was measured again.

\section{Results and discussion}

\begin{figure}
\includegraphics[width=8.5cm]{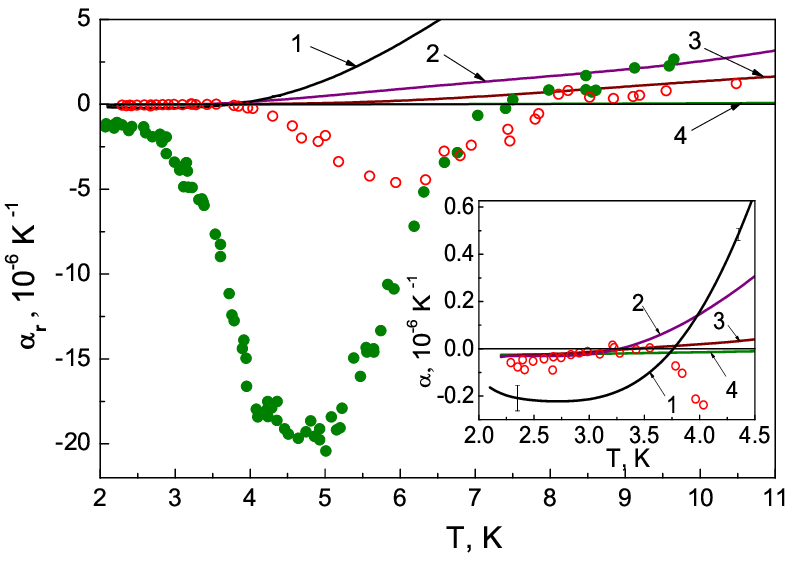}
a)
\includegraphics[width=8.5cm]{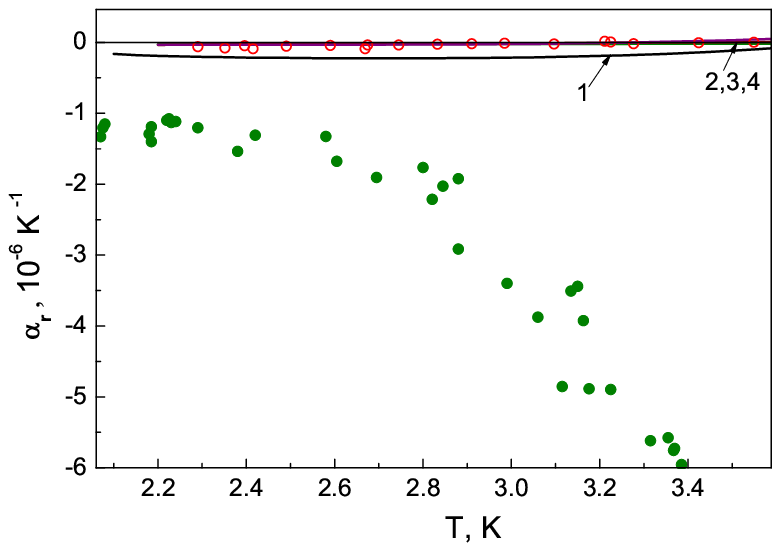}
b)
\caption{\label{fig:1}The radial thermal expansion coefficient 
$\alpha_r$ as function 
of temperature for SWNT bundles saturated with different gases.
(''b`` is a blowup of ''a``)\\*
Symbols show data from the present study: \textbullet{} -- $^3$He-SWNT, 
molar $^3$He concentration 9.4\%;
$\circ$ -- $^3$He-SWNT after partial removal of the $^3$He impurity at T~=~11~K.
Full lines show data from earlier work: 
1. $^4$He-SWNT, molar $^4$He concentration 9.4\% \cite{Dolbin-36-797-2010},
2. H$_2$-SWNT \cite{Dolbin-35-1209-2009}, 
3. Xe-SWNT \cite{Dolbin-35-613-2009}, and 
4. data for pure SWNTs \cite{Dolbin-34-860-2008}.
}
\end{figure}

The temperature dependence of the radial thermal expansion coefficient 
$\alpha_r$ of the $^3$He-SWNT system is shown in Fig.~\ref{fig:1}a. 
Solid circles represent $\alpha_r$ of the sample with the initial 
$^3$He concentration 9.4~mol.~\%, empty circles data for the same 
sample after partial removal of the $^3$He impurity by heating at 11~K. 
The inset in Fig.~\ref{fig:1}a shows the low-temperature data for the partially 
evacuated sample on an expanded scale to enable a comparison with 
earlier studies, while Fig.~\ref{fig:1}b is shown on an intermediate scale for
further comparisons with the saturated sample.

It is obvious from the Figure that saturating SWNT bundles with $^3$He 
causes a dramatic increase in the magnitude of the negative thermal 
expansion in the interval 2.1--7 K. The largest negative $\alpha_r$ 
in the 9.4\% $^3$He-CNT solution exceeds those of pure CNTs and 
$^4$He-saturated CNTs by three and two orders of magnitude, respectively. 
As in the case of the $^4$He-SWNT solution, the negative contribution 
to the thermal expansion of the $^3$He-SWNT system is most likely due to 
a process of spatial redistribution of the $^3$He atoms by tunneling 
at the surface and inside SWNT bundles. The isotope effect occurs 
because of the larger tunneling probability of the $^3$He atoms, 
which have a smaller mass than the $^4$He atoms, all other things being equal.

The temperature regions for the strong Schottky-like anomalies observed 
in the thermal expansion coefficients for the $^3$He-SWNT and $^4$He-SWNT 
solutions (similar anomailes have been predicted in specific heat
of low-density He gas adsorbed in carbon nanotube bundles 
\cite{Siber-66-075415-2002})
suggest rather low energy barriers impending the motion of the 
He atoms in SWNT bundles. Strzhemechny and Legchenkova \cite{Strzhemechny2011} 
have used the potential curves \cite{Firlej-241-149-2004} for a helium 
atom interacting with the outer surface of a single-wall carbon nanotube 
in order to evaluate the tunneling probability of different He isotopes 
along the nanotube. They showed that in this direction a $^4$He atom 
propagates in an energy band approximately 10.1~K wide. 
The respective band width for $^3$He is 13.4~K. 
These values are quite consistent with the results of this study.

After a partial $^3$He desorption from the sample the negative contribution 
of the impurity to the thermal expansion decreases and shifts towards higher 
temperatures (Fig.~\ref{fig:1}a). 
The reason may be as follows. There are several kinds of sites where He 
atoms can reside in SWNT bundles, and the resulting total tunneling 
contribution to the thermal expansion is a sum of contributions made 
by various types of tunneling motion. On desorption, the He atoms 
leave first the sites with a comparatively low energy for the bond 
between the He atom and the C framework. As a consequence, the role 
of different types of tunnel motion changes, which affects the temperature 
dependence of the resulting tunneling contribution to the thermal expansion.

The authors are indebted to M.~A.~Strzhemechny for fruitful 
discussions and to the National Academy of Sciences of Ukraine 
for the financial support of the study 
(Program ``Fundamental problems of nanostructural systems, 
nanomaterials, nanotechnologies'', Project ``The quantum phenomena 
in nanosystems and nanomaterials at low temperatures'').

\bibliography{article}

\end{document}